\documentstyle[11pt,newpasp,twoside,epsf]{article}
\reversemarginpar

\begin{document}
\title{Effects of the X-ray emission from young stars on the ionization
level of a fractal star forming cloud}
 \author{Andrea Lorenzani and Francesco Palla}
\affil{Osservatorio Astrofisico di Arcetri, Largo E. Fermi 5, 
50125 Firenze, Italy}

\begin{abstract}
We present the initial results of a study aimed at computing the global
effects of the X-ray ionization due to young stars within a molecular cloud 
that forms stars with a standard IMF at a rate accelerating with time.
We model the gas distribution of the cloud as a fractal of dimension 2.3.
We introduce the concept of a {\it R\"ontgen sphere} as the region around
each PMS star where the ionization rate due to stellar X-rays exceeds that
due to interstellar cosmic rays.  Using values of the X-ray luminosity
typical of T Tauri stars, we find that within the
R\"ontgen radius the average ionization rate is increased by a factor 4 to 20 
with respect to the value obtained from cosmic-rays.
\end{abstract}

\section{Introduction}
The initial results of the {\em Einstein Observatory}
have shown that young stars are powerful X-ray emitters
(Montmerle et al. 1983). It was then 
noted that the incident X-ray flux from YSOs may be the dominant
controlling agent of the cloud ionization (Krolik \& Kallman 1983).  
The possibility that X-ray ionization constitutes a natural feedback
mechanism in the physics of star formation was suggested by 
Silk \& Norman (1983). They argued that, in the framework of magnetically
controlled cloud evolution and collapse, the rate of star formation both 
determines, and is determined by, the cloud ionization level.
On a smaller scale, also the chemistry of the cloud around the immediate 
environs of a low-mass young star is deeply affected by the interaction
of the high-energy photons with the dense molecular gas (Glassgold, Najita 
\& Igea 1997).

The aim of this work is to re-examine the feedback mechanism of stellar
X-rays on ambient gas in light of current views of the history of star
formation in clusters and associations (e.g. Palla \& Stahler 2000) and of
the expanded knowledge of the X-ray properties of both embedded and optically
revealed young stars (Feigelson \& Montmerle 1999).  Observational studies in
a variety of star forming regions  have indicated that the X-ray activity
lasts longer and at higher levels than estimated in the initial studies.
Also, it appears that molecular clouds do not spawn stars at constant rates
during their lifetimes, but with a seemingly accelerating pattern. Finally,
observations of molecular clouds have shown that with increasing
spatial resolution the gas distribution breaks up into substructure 
of yet smaller scales.  This suggests that the density distribution  
may have a  fractal structure (e.g. Falgarone, Phillips \&
Walker 1991).

Obviously, all these properties can enhance the impact of the interaction
of the stellar X-rays with the surrounding medium.  
To assess such effects quantitatively, we have
first examined the distribution of X-ray energy around a star with a
characteristic X-ray luminosity $L_{\rm X}/L_{\rm bol} \la 10^{-3}$, typical
of T-Tauri stars.  We have then computed the overall contribution of the X-ray
ionization due to an ensemble of stars formed inside a molecular cloud of
sizes between 0.5 and 2 pc.  The basic question we want to
address is: {\it what fraction of the mass of the cloud is affected by the
enhancement in the ionization level due to X-rays?}

\section {The R\"ontgen Radius}
The ionization rate, $\zeta_{\rm X}$, due to X-rays from a YSO as
a function of  distance, $r$, from the star can be evaluated from the
expression:

\begin{equation}
\zeta_{\rm X} \ =\  1.7 \,\frac{L_{\rm X} \tilde{\sigma}}{4\pi r^2 \Delta \epsilon} \ 
\frac{\int_{\nu_0}^{\infty} J_{\nu}\, e^{- \tau_{X}(\frac{\nu}{\nu_{\rm X}} )^{-n}}
\big{(} \frac{\nu}{\nu_{\rm X}} \big{)}^{-n}  d \nu}{\int_{\nu_0}^{\infty} 
J_{\nu} \, d\nu} \,\ {\rm s}^{-1}
\end{equation} 
 
\noindent where $L_{\rm X}$ is the X-ray luminosity of the star,  
$\sigma(E) =\tilde{\sigma}(E/1 {\rm keV})^{-n}$ is the total photoelectric 
cross section with $n= 2.51$ and $\tilde{\sigma}=2.16\times 10^{-22}$ cm$^2$.
$\Delta \epsilon=35$ eV is the mean energy to make a ion pair in 
weakly-ionized gas with cosmic abundance, and the factor 1.7 accounts for the 
ionization due to secondary electrons (Cravens \& Dalgarno 1978). In eq. (1), 
$\tau_{\rm X}=n_{\rm H}\tilde{\sigma}r$ is the optical depth at $h\nu$=1 keV, 
$J_{\nu}$ is the monochromatic flux   emitted from the star, in the standard assumption
of a thermal bremsstrahlung spectrum. The minimum photon energy in the
integral of eq. (1) is $h\nu_0=$ 0.1 keV.
Assuming an X-ray luminosity of L$_{\rm X}=10^{30}$ erg s$^{-1}$,
the resulting run of the ionization rate with distance from
the central star is displayed in Figure 1.

\begin{figure}[h]
\plotfiddle{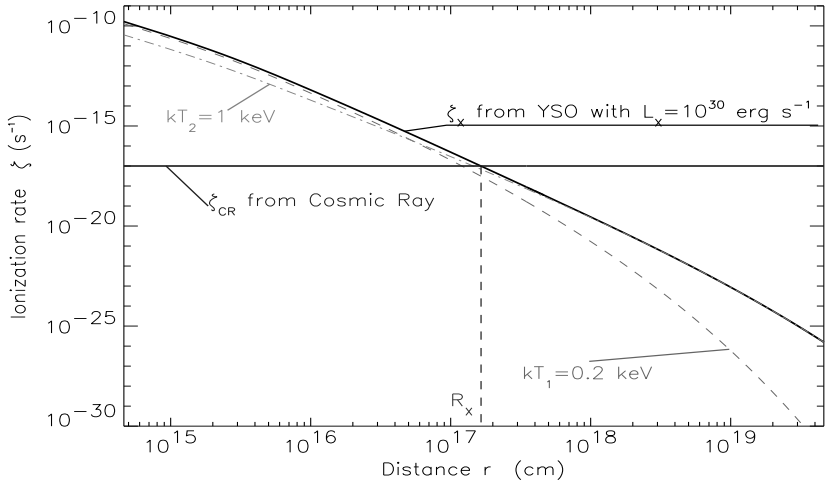}{6cm}{0}{65}{55}{-160}{-7}
\caption{The X-ray ionization rate versus distance from the star, for a
YSO with $L_{\rm X}=10^{30}$ erg s$^{-1}$ and a two temperature spectrum
model. $R_{\rm X}$ is the R\"ontgen radius. }
\end{figure}

As we see, interior to a radius $R_{\rm X}$ the ionization rate due to X-rays
exceeds the background level provided by cosmic rays,
here assumed at the constant value $\zeta_{\rm CR}=10^{-17}$ s$^{-1}$.
We call this X-ray dominated region the {\em R\"ontgen sphere}, 
and $R_{\rm X}$ the {\em R\"ontgen radius}. In the case of Fig. 1,
the average ionization rate inside the sphere of constant density
n$_{\rm H}=10^4$ cm$^{-3}$ is about 20 times that due to cosmic rays.
We note that the uncertainty on the empirical value of $\zeta_{CR}$ is 
a factor of about five around our assumed value (e.g. van der Tak \& van
Dishoeck 2000).

To gauge the dependence of $R_{\rm X}$ and $\zeta_{\rm X}$ on the model
parameters, we have considered X-ray emission spectra at different plasma
temperatures, as well as the effects of an absorbing column density of N$_{\rm
H}=5\times10^{20}$ cm$ ^{-2}$ in front of the X-ray emitting region.  As
shown in Table~1, the  R\"ontgen radius is insensitive to the input values,
while the average ionization rate can vary by a factor of $\sim$3 (the label
$w$ refers to the model with local absorption).

\begin{table}[h]
\begin{center}
\caption {R\"ontgen radius and average ionization rate}
\begin{tabular}{l c| c c| c c} 

\tableline
Plasma & kT & $R_{{\rm X}w}$ & $\zeta_{{\rm X}w}$ &  $R_{\rm X}$ &   $\zeta_{\rm X}$  \\
model &(keV)&(10$^{17}$ cm)&(10$^{-17}$ s$^{-1}$)&(10$^{17}$ cm)& 
(10$^{17}$ s$^{-1}$) \\
\tableline

2T\ & 0.2\  +\  1  & 1.5 & 6.3 & 1.6 & 21.0 \\
1T\ & 1            & 1.4 & 5.1 & 1.5 & 10.5\\
1T\ & 3            & 1.1 & 4.3 & 1.2 & 7.8\\
\tableline
\tableline
\end{tabular}
\end{center}
\end{table}

\section {Star formation within a fractal cloud}

The results of the previous section for an individual star can now be extended
to an ensemble of young stars formed in a cluster/association. The assumptions
that enter in the model are:

\begin{itemize}

\item The dense gas distribution has a fractal structure of dimension 2.3
(Falgarone, Phillips \& Walker 1991). A fractal cloud model is obtained  from
hierarchically clustered points. Each point is divided into N=5 random
points, all within a distance $1/L^h$ for hierarchical level $h$ and
geometric factor $L=2$ (Elmegreen 1997). The fractal dimension is defined as
$D=\log N/ \log L \sim 2.3$. In Fig.~2 we display a rendition of the fractal
gas distribution, integrated along the line of sight. Here, the gas density
varies from 500 cm$^{-3}$ to 10$^6$ cm$^{-3}$ and the smallest structure is
2$^{-6}$ times the characteristic size of the cloud ($\sim$1.5 pc).

\begin{figure}[h]
\plotfiddle{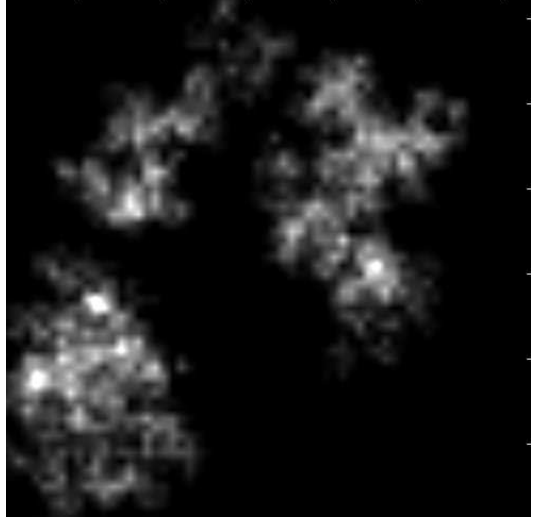}{6cm}{0}{40}{40}{-110}{-60}
\caption{Bidimensional representation of the fractal cloud with D=2.3.}
\end{figure}

\item We assume that at $t=0$ the cloud contains no stars. Star formation
then begins and continues for 10 million years. Initially, the rate $\Psi(t)$
is low and then accelerates exponentially as $\Psi(t) \propto e^{\Delta
t/t_{\rm c}}$, with $t_{\rm c}=2$~Myr (Palla \& Stahler 2000).

\item At any time, star are formed randomly in mass in the interval 0.1--10
M$_\odot$.  However, we require that after 10 million years their
distribution $\phi(m)$ follows a standard IMF $\phi(m)\sim m^{-\gamma}$, with
$\gamma$ given by, e.g., Scalo (1998).  The behavior of both the star
formation rate (SFR) and the mass distribution with time is displayed in the
left panel of Figure~3.

\begin{figure}[h]
\plottwo{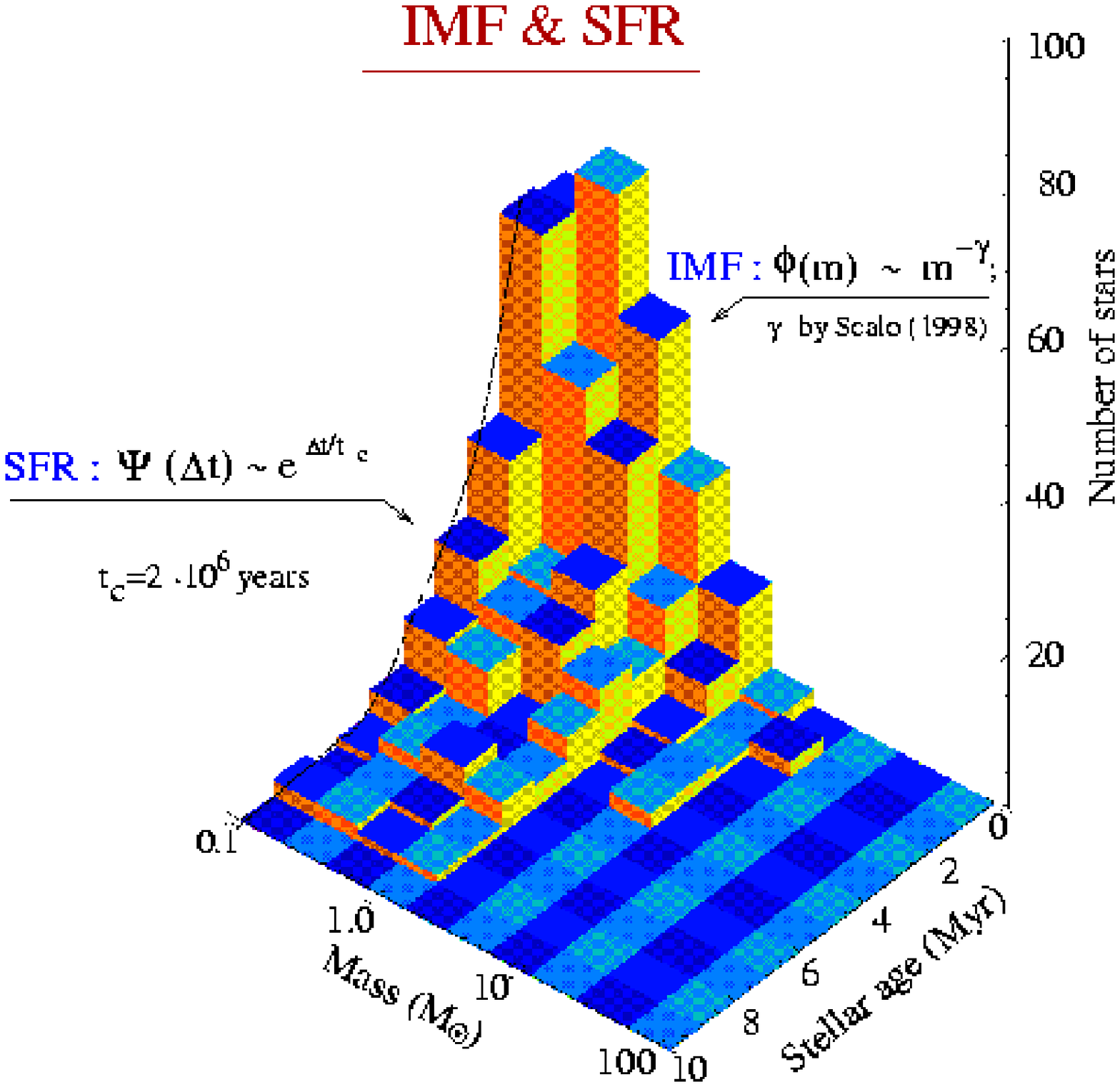}{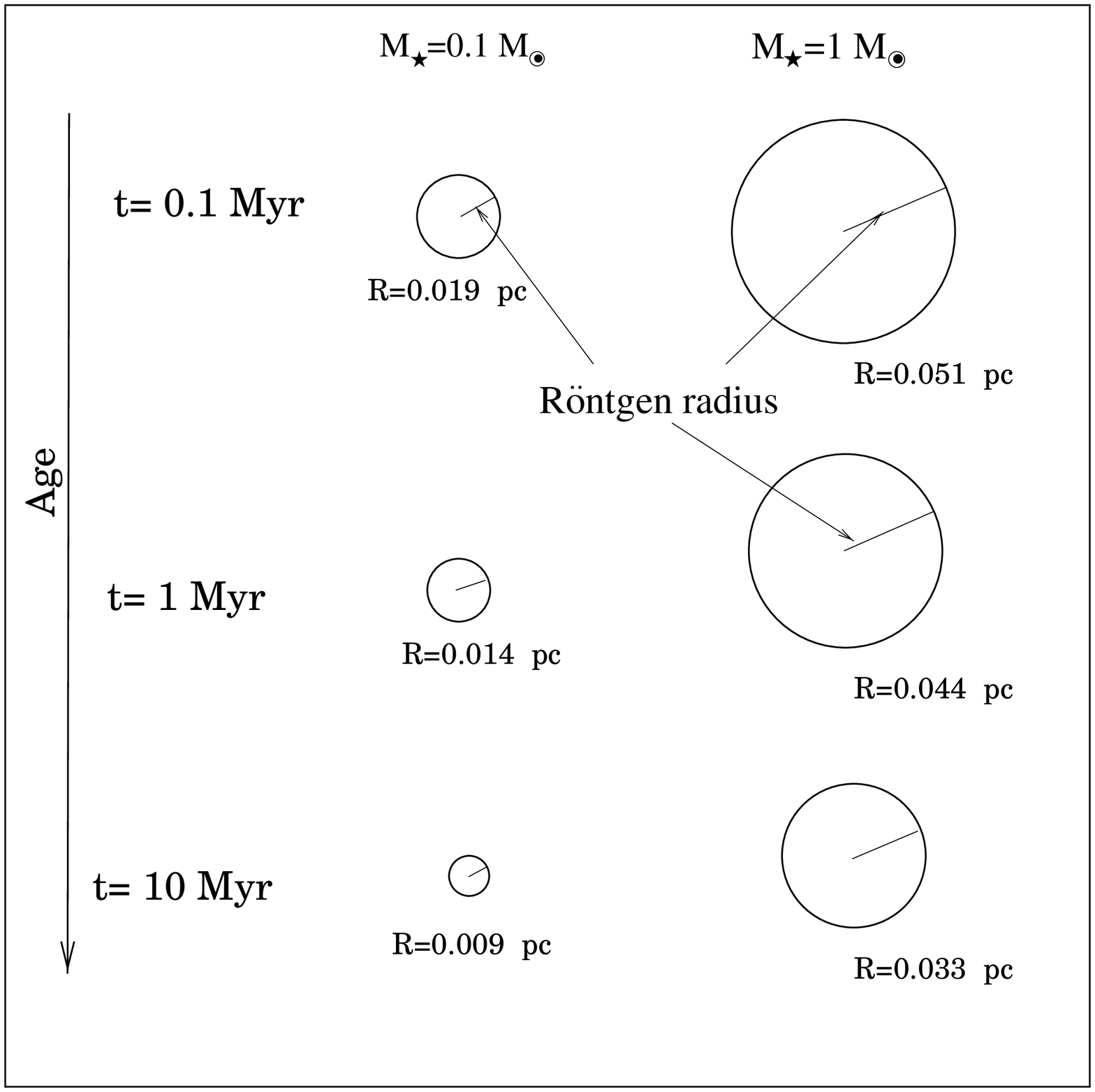}
\caption{
{\it Left panel}: Time evolution of the SFR and mass distribution of 500 
stars born in 10 Myr. 
{\it Right panel}: The evolution of the R\"ontgen radius in an ambient medium
of 10$^4$ cm$^{-3}$ during the PMS contraction
of two stars of mass $0.1 {\rm M}_{\sun}$ and $1 {\rm M}_{\sun}$, respectively. 
}
\end{figure}

\item  We assume that the stellar X-ray luminosity is proportional to its
bolometric luminosity:  $L_{\rm X}(M_{\star},t)=3\times 10^{-4} {\rm L}_{\rm
bol}({\rm M}_{\star},t)$.  Each star has its associated R\"ontgen sphere,
computed as in Sect. 2.  Because of the variation of ${\rm L}_{\rm bol}({\rm
M}_{\star},t)$ during PMS contraction, the R\"ontgen radius also varies with
time, as illustrated in Fig.~3 (right panel).

\end{itemize}

\section{X-ray ionization of a fractal cloud}

Figure~4 shows a 3-D view of the evolution of a fractal cloud
of size 1.5 pc following the prescriptions given above.
Four different epochs are selected: regions coded in light gray represent
the individual R\"ontgen spheres formed randomly within the cloud.
After 10 million years,
the cumulative X-ray luminosity of all the stars formed in the cloud 
is $6\times 10^{32}$ erg s$^{-1}$, 
and the overall star formation efficiency is about 16\%.  

\begin{figure}[h]
\plotfiddle{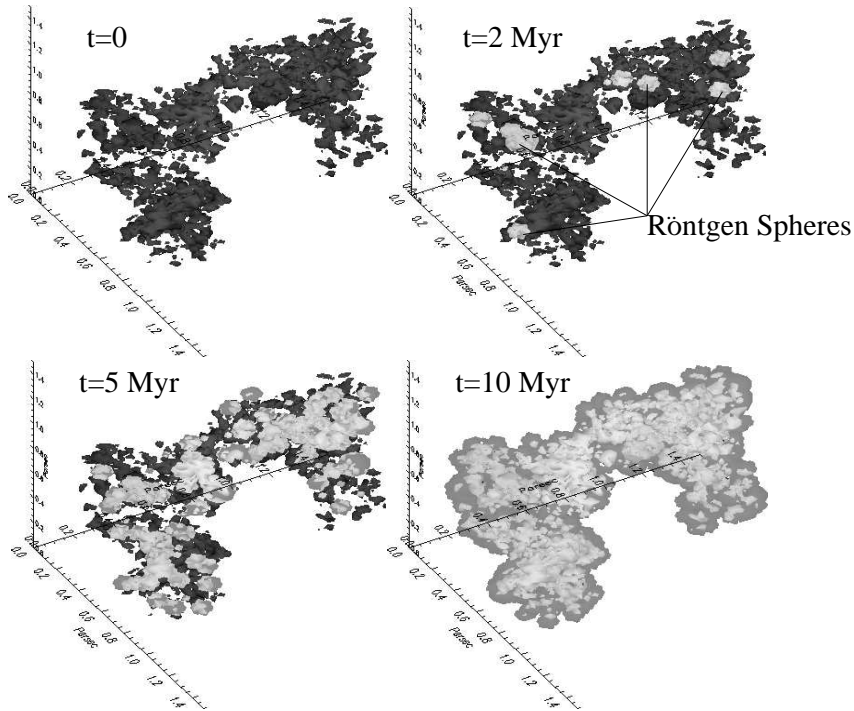}{8.5cm}{0}{50}{50}{-160}{-10}
\caption{Time evolution of the R\"ontgen spheres in the fractal cloud.}
\end{figure}

In a fractal cloud of size $R_0$ the mass enclosed in a radius R$_{\rm X}$ is
proportional to $(R_{\rm X}/R_0)^D$, where $D$ is the fractal dimension.
Thus, the mass fraction, $F$, of the cloud ionized by the stellar X-rays at
time $t_{\rm f}=10^7$~yr can be computed as:

\begin{equation}
F(t_{\rm f})=\int_{m_{\rm min}}^{m_{\rm max}}\int_{0}^{t_{\rm f}}\phi(m) \Psi(t) (R_{\rm X}(m,t_{\rm f}-t)/
R_0)^D\ dm\ dt
\end{equation}
\noindent where  ($m_{\rm min},m_{\rm max}$) is the range of stellar masses considered.
Thus, we can estimate the total number of stars needed to have $F(t_{\rm f})=1$, 
i.e. complete overlap of the R\"ontgen spheres.
In Table~2 we report these values for clouds with different sizes, along
with the relative stellar density.

\begin{table}[h]
\caption{Variation of the number of stars for $F=$1}
\begin{center}
\begin{tabular}{c | c c | c c}

\tableline
 Cloud Size & N. stars & $\rho_\star$ & N. stars$_w$ & $\rho_{\star w}$ \\
(pc) &  & (pc$^{-3}$) & & (pc$^{-3}$) \\
\tableline
0.5 &90 &700 & 100& 800  \\
1.0 &400&400 & 500& 500 \\
1.5 &800&250 & 1200& 350 \\
2.0 &1500&190& 2000& 250 \\
\tableline
\tableline
\end{tabular}
\end{center}
\end{table}

\section{Conclusions}

The preliminary results described above indicate that the X-rays produced
by young stars in a cluster forming cloud may contribute significantly
to its ionization degree, especially if the dense gas is not distributed
uniformly. Typical R\"ontgen radii vary between
$\sim 0.01-0.05$ pc, with an enhancement of the ionization rate
of a factor $ \simeq\ 5-20$ over the value due to cosmic-rays alone. 
The mass fraction of a molecular cloud ionized by X-rays and
cosmic rays is a function of the total number of stars and of their X-ray
luminosity. For a fixed 
X-ray luminosity of $L_{\rm X} \sim 3\times 10^{-4} L_{\star}$, the stellar
density needed to fill a cloud of 1 pc size
is about $\rho_\star \geq 500$~pc$^{-3}$. The
required number of stars,  the total X-ray luminosity and the formation
efficiency are thus comparable to those observed in cluster forming regions,
such as $\rho$ Ophiuchi.
The ionization due to YSOs may therefore increase the coupling of the magnetic field
to the cloud and the characteristic time of cloud core collapse via
ambipolar diffusion. If   star formation  is indeed {\em accelerating},
the X-ray emission from YSOs can constitute a natural feedback mechanism to
{\em decelerate} the process. More realistic models (including the effects
 of the dynamical evolution of the gas during star formation, a distribution
 of X-ray
luminosity and emission spectrum as a function of mass and age, etc.) are
being carried out to verify the  initial results.

\end{document}